\documentclass[letterpaper,english,aps,pra,twocolumn,amsfonts, amssymb]{revtex4-2}
\usepackage[T1]{fontenc}
\usepackage[utf8]{inputenc}
\usepackage{textcomp}
\usepackage{mathrsfs}
\usepackage{amsmath}
\usepackage{amssymb}
\usepackage{amsfonts}
\usepackage{makeidx}
\usepackage{graphicx}
\usepackage{esint}
\usepackage[normalem]{ulem}

\makeatletter
\usepackage[bookmarks=true,colorlinks,linkcolor=blue,urlcolor=blue,citecolor=blue]{hyperref}
\usepackage{physics}
\newcommand{\be}{\begin{equation}}
\newcommand{\ee}{\end{equation}}
\newcommand{\bea}{\begin{eqnarray}}
\newcommand{\eea}{\end{eqnarray}}

\usepackage{epsfig,graphicx,psfrag,amsmath,amssymb,float}
\usepackage[caption=false]{subfig}
\@ifundefined{showcaptionsetup}{}{%
\PassOptionsToPackage{caption=false}{subfig}}
\usepackage{subfig}
\makeatother
\usepackage{babel}
\usepackage{float}

\begin{document}

\title{Non-Hermitian dynamics and $\mathcal{PT}$-symmetry breaking in interacting mesoscopic Rydberg platforms}

\author{Jos\'e A. S. Louren\c{c}o$^{1,2}$}

\author{Gerard Higgins$^2$}

\author{Chi Zhang$^2$}

\author{Markus Hennrich$^2$}

\author{Tommaso Macrì$^1$}

\affiliation{
	$^1$Departamento de F\'isica Te\'orica e Experimental, Universidade Federal do Rio Grande do Norte, Campus Universit\'ario, Lagoa Nova, Natal, RN, 59072-970, Brazil \\
	$^2$Department of Physics, Stockholm University, 10691 Stockholm, Sweden}

\begin{abstract}
We simulate the dissipative dynamics of a mesoscopic system of long-range interacting particles which can be mapped into non-Hermitian spin models with a $\mathcal{PT}$ symmetry. We find rich $\mathcal{PT}$-phase diagrams with $\mathcal{PT}$-symmetric and $\mathcal{PT}$-broken phases.
The dynamical regimes can be further enriched by modulating tunable parameters of the system.
We outline how the $\mathcal{PT}$ symmetries of such systems may be probed by studying their dynamics.
We note that systems of Rydberg atoms and systems of Rydberg ions with strong dipolar interactions are particularly well suited for such studies.
We show that for realistic parameters, long-range interactions allow the emergence of new $\mathcal{PT}$-symmetric regions, generating new $\mathcal{PT}$-phase transitions. In addition, such $\mathcal{PT}$-symmetry phase transitions are found by changing the Rydberg atoms configurations. We monitor the transitions by accessing the populations of the Rydberg states. Their dynamics display oscillatory or exponential dependence in each phase.
\end{abstract}

\maketitle


\section{Introduction}

Closed quantum systems evolve with unitary dynamics according to the Schrödinger equation; the Hamiltonian of such a system is Hermitian and the eigenenergies are real.
A dissipative quantum system evolves with non-unitary dynamics, which, in some cases, can be approximately described according to a non-Hermitian Hamiltonian \cite{moiseyev2011,Rotter,Eckardt_2015}. Some of these systems are invariant under the simultaneous application of parity ($\mathcal{P}$) and time reversal ($\mathcal{T}$) symmetry operators \cite{Bender1998}.
If all the eigenenergies of the non-Hermitian Hamiltonian are real, a system is said to preserve $\mathcal{PT}$-symmetry, otherwise if part of the spectrum is complex $\mathcal{PT}$ symmetry will be broken \cite{Bender2}.
The critical values that limit the $\mathcal{PT}$-unbroken and $\mathcal{PT}$-broken phases are the exceptional points \cite{Heiss2012,Wei2017}. Also, the existence of a continuous set of exceptional points that limit regions with $\mathcal{PT}$-unbroken and $\mathcal{PT}$-broken phases are called exceptional lines \cite{Moors2019,Yang2019,Zhang2020prb}.
$\mathcal{PT}$-symmetry of several dissipative quantum systems has been studied in different contexts, e.g. in optics \cite{Ruter2010,Zyablovsky_2014,Lodahl2017,Ramy2018, Longhi:181,*Longhi:182,*Longhi:183,*Longhi:19,*Longhi:20,Benderbook2019}, photonics \cite{Feng2017,Zhao2018,Ozdemir2019,Klauck2019}, quantum many-body systems \cite{Fei2003,Korff2007,Olalla2009,Deguchi2009,Bender_2015,Ashida2016,*Ashida2017,*Ashida2018,*Ashida2020,Wei2017,Kawabata2017,Lourenco2018,Nakagawa2018,Hamazaki2019,Yamamoto2019,Xiao2019,Matsumoto2019,Lee2020,Shackleton2020,Weidemann2020,Takasu2020,Julian2020,Yamamoto2020, Diego2021}, systems with topological models \cite{Yuce2018,*Yuce2018.2,Wang2019,Longhi2019.6,Kawabata2019topol,Kawasaki2020,Chang2020,Fate2020,Blose2020,Guo_2020,Mittal2020,Xia2020} and in curved space \cite{Ygor2021}.
The $\mathcal{PT}$ transition was verified experimentally for a cold atomic dissipative Floquet system, in which the $\mathcal{PT}$ symmetry transitions can occur by tuning either the dissipation strength or the coupling strength. \cite{Li2019}. Recently, models with Floquet $\mathcal{PT}$-symmetric modulation showed a $\mathcal{PT}$ transition in square wave modulation \cite{2020Andrew1,2020Andrew2}, and were experimentally demonstrated in a system of noninteracting cold fermions \cite{Li2019}. Experimentally, a single trapped ion was used to investigate the dynamics of $\mathcal{PT}$-symmetric non-Hermitian systems \cite{Wang2021,Ding2021}.

In this work we investigate the non-Hermitian dynamics of a mesoscopic system of few particles coupled via long-range interactions among them \cite{Defenu2021}. 
We find that the interactions and geometric arrangement of coupled spins enrich the $\mathcal{PT}$ phase diagrams (Sec.~\ref{sec2}).
We also find that by modulating system parameters the $\mathcal{PT}$ nonequilibrium dynamical phase diagrams can be further enriched (Sec.~\ref{sec2b}).
The $\mathcal{PT}$ phase can be determined by probing the system's dynamics, and reconstructing the system's eigenenergies (Sec.~\ref{sec3}).
Our analysis is inspired by recent experiments with ultracold Rydberg atoms and ions, which are well suited for experimentally probing these effects in (Sec.~\ref{sec3a}). Furthermore, our model can be generalized to more complete Rydberg atom models and more complex systems \cite{Beguin2013,Macri2014,Barredo2015,Zeiher2015,Labuhn2016,Ravets:16,Browaeys2016Experimental,*Browaeys:hal-01717168,*Browaeys2020Many,Scholl2021Quantum,Sylvain2019Observation,Henriet2020quantumcomputing,Hermes2020}, and can also be implemented on other experimental platforms \cite{Yan2020resonant,Lin2020Quantum}.

\begin{figure}[ht!]
	\centering
	\includegraphics[height=5cm]{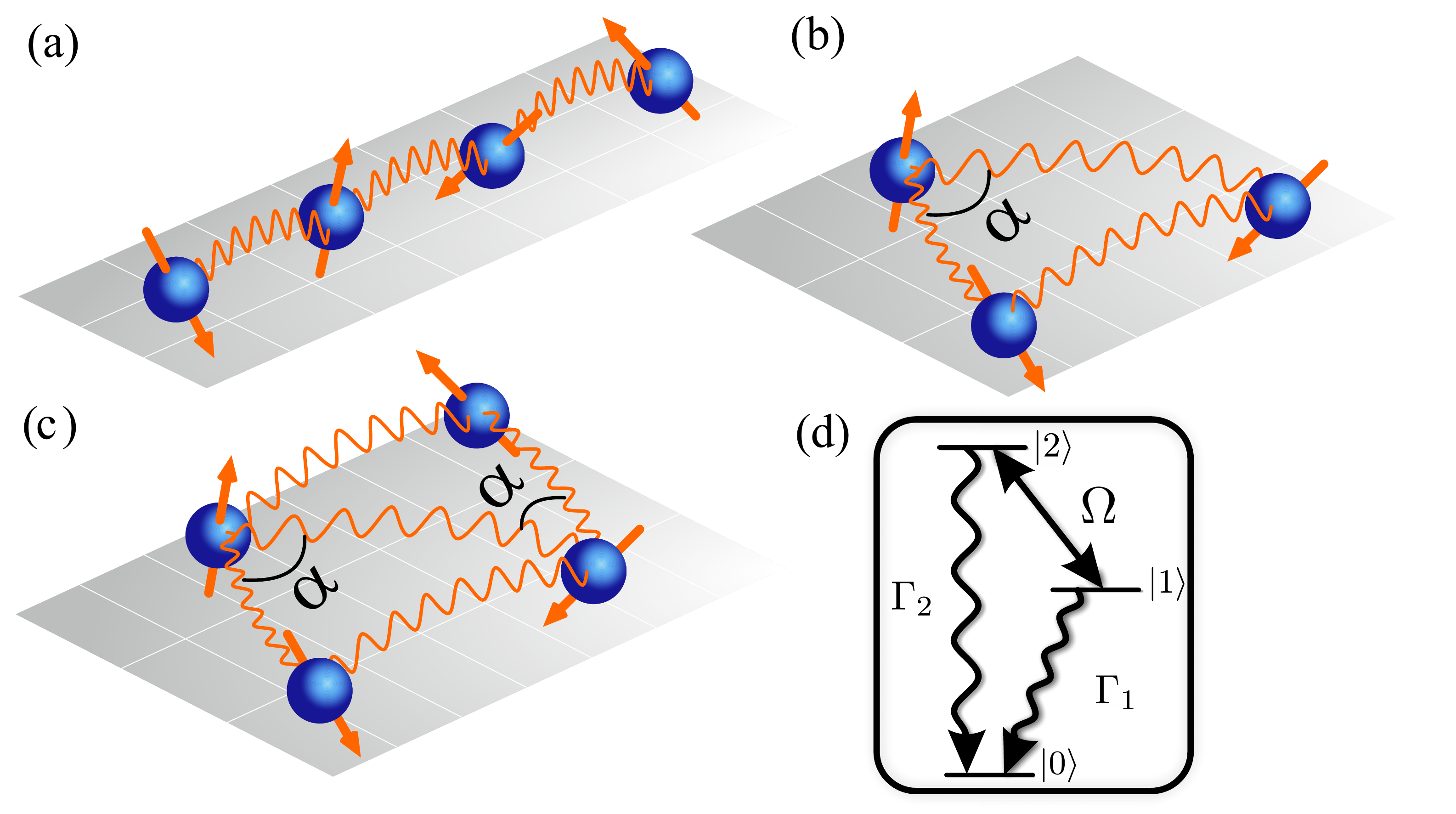}
	\label{fig5}
	\caption{\label{fig1} (a) Linear configuration of interacting 4-particles. Zigzag configurations parameterised by angle $\alpha$: (b) three-particle zigzag and (c) four-particle zigzag. (d) Level scheme of the internal structure of the particles leading to an effective non-Hermitian dynamics. Each particle has two levels $\ket{1}$ and $\ket{2}$ which are coupled with strength $\Omega$, and which decay to a third state $\ket{0}$ with rates $\Gamma_1$ and $\Gamma_2$ respectively.}
\end{figure}

\section{Model}\label{sec2}
We consider a system of $N$ particles modeled by two internal levels and loss channel to a third state. Particles interact via long-range exchange interactions.
We are motivated by recent works with strongly-interacting Rydberg ions \cite{Zhang2020}, and our model was designed to match this system.
In Sec.~\ref{sec3a} we describe in more detail about how this model can be experimentally implemented using Rydberg ions.

The two levels of each particle are labeled by $\ket{1}$ and $\ket{2}$. A third level $\ket{0}$ represents all the states to which $\ket{1}$ and $\ket{2}$ can decay, with decay rates $\Gamma_{1}$ and $\Gamma_{2}$ respectively. States $\ket{1}$ and  $\ket{2}$ are Rabi-coupled with coupling strength $\Omega$,  as shown in Fig.$~$(\ref{fig1}). 
The $i^\mathrm{th}$ and $j^\mathrm{th}$ particles interact with strength $V_{ij}$, which can be assumed to be a e.g. dipolar ($\propto |i-j|^{-3}$) interaction.
When written in terms of spin operators, the interaction takes the form of an $XY$ exchange interaction between particles $i$ and $j$.

The system is described by the effective non-Hermitian Hamiltonian
\begin{eqnarray}\label{eq2}\nonumber
	\hat{H}_{{\rm eff}}&=&\sum^{N}_{i=1}\frac{1}{2}\left[\Omega\hat{\sigma}_{x}^{i}-i\Gamma\hat{\sigma}_{z}^{i} -i\frac{\Gamma_{1}+\Gamma_{2}}{2}\hat{\mathbb{I}}^i\right]\\
	&&+\sum^{N}_{i>j} \frac{V_{ij}}{2}\left(\hat{\sigma}_{x}^i\hat{\sigma}_{x}^j+\hat{\sigma}_{y}^i\hat{\sigma}_{y}^j\right).
\end{eqnarray}
where $\mathbb{\hat{I}}^i$ is the identity operator and $\Gamma \equiv \left(\Gamma_1 -\Gamma_2\right)/2$. The operators $\sigma_{z} = \ketbra{2}{2}-\ketbra{1}{1}$, $\hat{\sigma}_x = \hat{\sigma}_{+} + \hat{\sigma}_{-}$, $ \hat{\sigma}_y = -i\left(\hat{\sigma}_{+} - \hat{\sigma}_{-}\right)$, and the raising and lowering operators are $\hat{\sigma}_{+} = \ketbra{2}{1}$ and $\hat{\sigma}_{-} = \ketbra{1}{2}$, respectively. The term in Eq.$~$(\ref{eq2}) proportional to $\mathbb{\hat{I}}$ does not affect the dynamics (it can be absorbed in the normalization of the state), 
and so we can write the $\mathcal{PT}$-symmetric non-Hermitian Hamiltonian as
\begin{equation} \label{eq3}
	\hat{\mathcal{H}}_{\mathcal{PT}}=\sum_{i=1}^{N}\frac{1}{2}\left(\Omega\hat{\sigma}_{x}^{i}-i\Gamma\hat{\sigma}_{z}^{i}\right) +\sum^{N}_{i>j} \frac{V_{ij}}{2}\left(\hat{\sigma}_{x}^i\hat{\sigma}_{x}^j+\hat{\sigma}_{y}^i\hat{\sigma}_{y}^j\right).
\end{equation}
Using the parity operator, $\mathcal{P}=\hat{\sigma}_{x}$, and the time reversal operator, $\mathcal{T}=K$, where $K$ is the complex conjugate operator, one can show that $\hat{H}_{\mathcal{PT}}$ in Eq.$~$(\ref{eq3}) is $\mathcal{PT}$-symmetric, i.e. $\left(\mathcal{PT}\right)\hat{H}_{\mathcal{PT}}\left(\mathcal{PT}\right)^{-1}=\hat{H}_{\mathcal{PT}}$.
Throughout this work we determine whether $\mathcal{PT}$ is preserved or broken from the eigenvalues of $\hat H_{\mathcal{PT}}$: When $\hat H_{\mathcal{PT}}$ has real eigenvalues the $\mathcal{PT}$ symmetry is preserved, otherwise the $\mathcal{PT}$ symmetry is broken.
Whether $\mathcal{PT}$ symmetry is preserved or broken depends on the relative strengths of the dissipative part of the Hamiltonian ($\Gamma$) and the non-dissipative part of the Hamiltonian ($\Omega$ and the interactions $V_{ij}$).

In the following sections we consider the $\mathcal{PT}$ symmetry phase diagrams of different configurations of interacting particles and when the system parameters are modulated.

\begin{figure*}
	\centering
	\includegraphics[height=18cm]{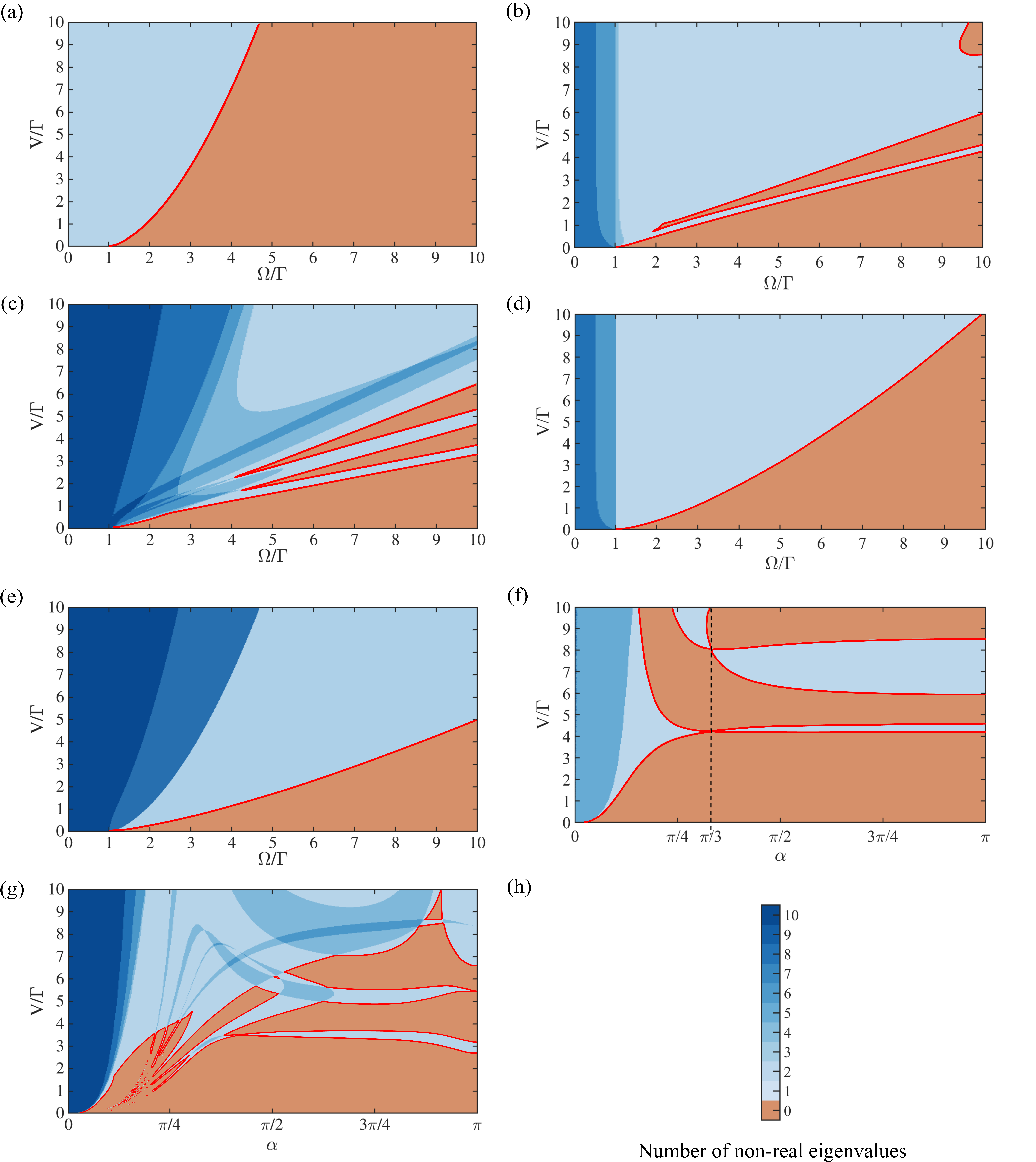}
	\caption{\label{fig3} $\mathcal{PT}$-symmetry phase diagrams for interacting particle systems. The symmetry of $\mathcal{PT}$ is preserved when all eigenvalues are real (orange regions), otherwise it is broken (shaded blue regions). $\mathcal{PT}$-phase transitions occur along exceptional lines (red lines). (a), (b) and (c) show results for $N=2$, $3$ and $4$ particles in the linear configuration; the diagrams get richer to increase the number of particles. (d) and (e) show results for particles in a triangular and tetrahedral configuration, respectively. (f) and (g) show results for $N=3$ and $N=4$ particles in zigzag configurations. In (f) the zigzag configuration becomes an equilateral triangle at $\alpha=\pi/3$, and the symmetry of the system is increased. In (f) and (g) we set $\Omega/\Gamma=10$.
	(h) Color bar with the number of non-real eigenvalues.}
\end{figure*}
\subsection{Different configurations of interacting particles}

\subsubsection{Single particle}
For a single-particle system ($N=1$), the eigenvalues of $\hat{H}_{\mathcal{PT}}$ are $\lambda_\pm=\pm\Gamma/2\sqrt{\left(\Omega/\Gamma\right)^{2}-1}$. The $\mathcal{PT}$-symmetry is preserved when $\Omega/\Gamma>1$ and the eigenvalues are real. The $\mathcal{PT}$-symmetry is broken when $\Omega/\Gamma<1$ and the eigenvalues are imaginary. The exceptional point occurs at $\Omega/\Gamma=1$ and the eigenvalues coalesce and $\lambda_\pm=0$. 
A system of non-interacting single particles was experimentally studied using ultra-cold atoms in \cite{Li2019} and for a single ion in \cite{Ding2021, Wang2021}.

In the rest of this work, we calculate the eigenvalues of $\hat{H}_{\mathcal{PT}}$ (and thus its $\mathcal{PT}$ symmetry) numerically for different experimentally relevant configurations with realistic interactions.

\subsubsection{Linear chain of particles}
Next we consider a equidistant linear chain of particles, in which the interaction strength falls with the cube of the separation $V_{ij}=V/|i-j|^3$, such as a dipolar interaction. 
We numerically calculated the eigenvalues of $\hat H_{\mathcal{PT}}$ for systems with two, three and four particles and found the $\mathcal{PT}$-symmetry of $\hat{H}_{\mathcal{PT}}$ depends on the relative strengths of $\Omega$, $\Gamma$ and $V$, as shown in Fig.~\ref{fig3}(a-c).

The ${\mathcal{PT}}$ symmetry phase diagrams become richer with increasing $N$.
While the $N=2$ case shows a single $\mathcal{PT}$-symmetry preserving region, more regions are present in the $N=3$ and $N=4$ cases. $\mathcal{PT}$ symmetry is preserved when all eigenvalues are real (orange regions), otherwise it is broken (other regions). $\mathcal{PT}$-phase transitions occur along exceptional lines (red lines) Note that when $V=0$, the exceptional point at $\Omega/\Gamma = 1$ (for the single-atom case) is recovered.

\subsubsection{Equilateral triangular and tetrahedral configurations}
Next we consider three particles arranged in an equilateral triangle configuration and four particles arranged in a tetrahedral configuration. We have $V_{ij}=V$ for each of the particle pairs.
We determine whether or not $\hat H_{\mathcal{PT}}$ is ${\mathcal{PT}}$-symmetry preserving by numerically calculating its eigenvalues as the parameters $\Omega$, $\Gamma$ and $V$ are varied; the ${\mathcal{PT}}$-phase diagrams are shown in Figs.$~$\ref{fig3}(d) and (e). 
These phase diagrams are less rich than the diagrams for the linear chains (see Figs.$~$\ref{fig3}(b) and (c)). We argue that the symmetry of the equilateral triangle and tetrahedral configurations constrains the dynamics, and leads to a single ${\mathcal{PT}}$-symmetry region.


\subsubsection{Zigzag configurations}
Next we consider $3$-particle and $4$-particle zigzag configurations, parameterised by the angle $\alpha$ as shown in Figs.$~$\ref{fig5}(a) and (b); the $\mathcal{PT}$-phase diagrams are presented in Figs.$~$\ref{fig3}(f) and (g).
As in the previous subsection, the $i^\mathrm{th}$ and $j^\mathrm{th}$ atoms interact with dipolar interactions.
We define the interaction scale $V_{12}=V$ between nearest neighbors; the interaction strengths between other atom pairs follows from geometrical arguments and are explicitly described in Appendix~(\ref{appendix_1}).
In computing the eigenvalues of $H_{\mathcal{PT}}$, we varied $V$ and $\alpha$, and we set $\Omega/\Gamma=10$.

The $\mathcal{PT}$-phase diagrams depend on the angle $\alpha$ as a result of the interaction strengths $V_{ij}$ dependence on $\alpha$.
We note that in the $3$-particle configuration when $\alpha=\frac{\pi}{3}$ (black dashed line in Fig.$~$\ref{fig3}(f)) the configuration becomes an equilateral triangle, and we posit that the increased symmetry of the configuration is linked to the crossovers of exceptional lines. In the case of the $4$-particle configuration, when $\alpha = \frac{\pi}{4}$ in Fig.$~$\ref{fig5}(b), we find the square configuration. The $\mathcal{PT}$-phase transitions of the $N=3,\,4$ particles linear configurations are retrieved in the limit $\alpha \rightarrow \pi$.

\subsection{System with time-dependent parameters}\label{sec2b}

The $\mathcal{PT}$-symmetry phase diagrams can be enriched further by periodically modulating the parameters $\Omega$, $V$ or $\Gamma$ with a charcteristic frequency $\omega_F$ \cite{Li2019}. When the parameters are modulated, the eigenvalues of $\hat H_{\mathcal{PT}}$ become time dependent.
Then, to study $\mathcal{PT}$ symmetry, instead of calculating whether the eigenvalues of $\hat H_{\mathcal{PT}}$ are real, we use the eigenvalues of the time-evolution operator
\begin{equation}\label{eq26.1}
\hat G_{\mathcal{PT}}(t) =\text{e}^{-i\hat H_{\mathcal{PT}}t}.
\end{equation}
When all the eigenvalues of $\hat G_{\mathcal{PT}}(\frac{2\pi}{\omega_\mathrm{F}})$ have unitary modulus, $\mathcal{PT}$ symmetry is preserved, otherwise it is broken. 

\begin{figure}[ht!]
	\centering
	\includegraphics[height=13cm]{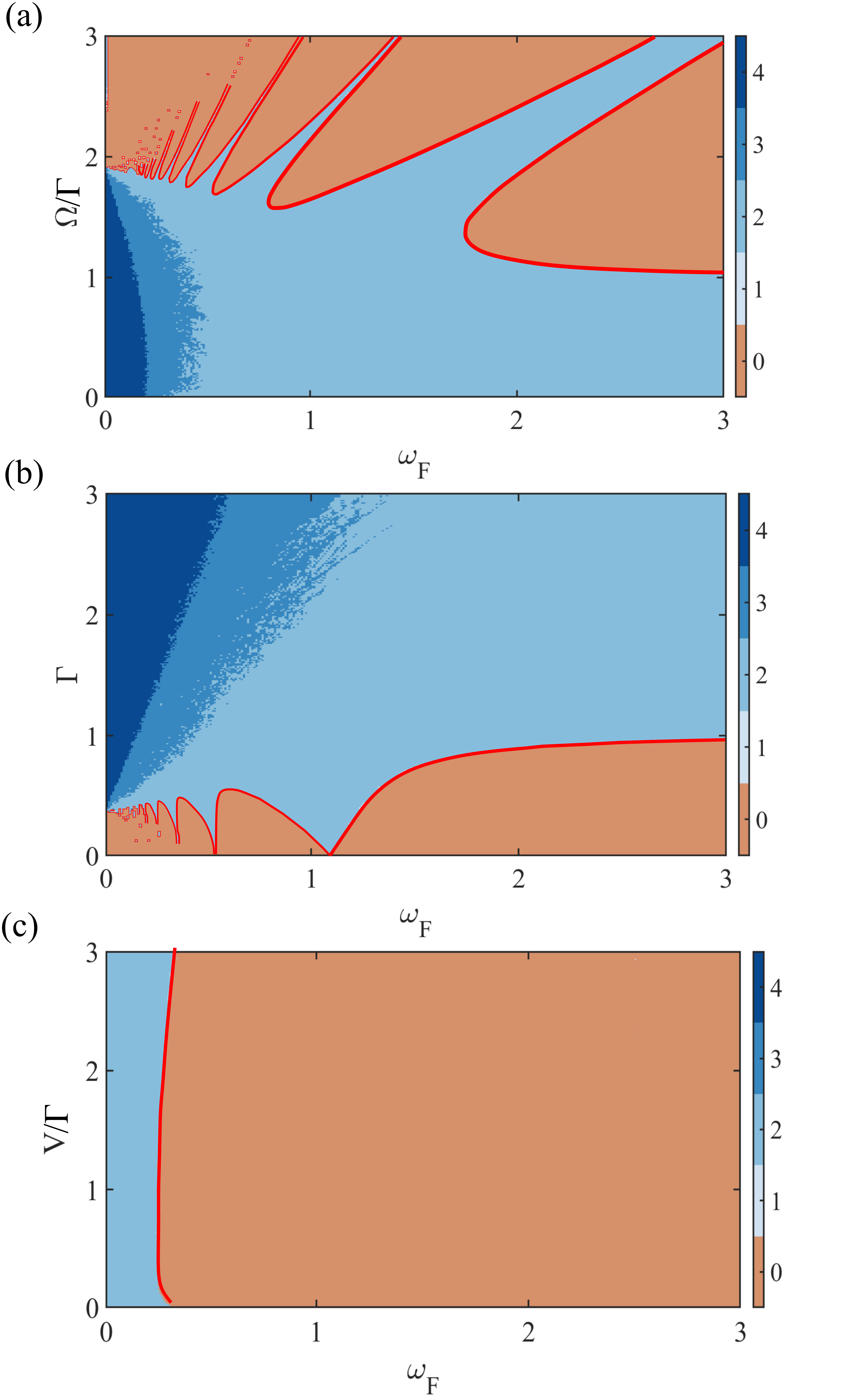}
	\caption{\label{fig12} $\mathcal{PT}$-phase diagrams for two-particle systems when $V(t)$ is sinusoidally modulated at frequency $\omega_\mathrm{F}$. $\mathcal{PT}$ symmetry is preserved when all the eigenvalues of the time-evolution operator have unit modulus (orange regions). (a) $\Omega \times \omega_\mathrm{F}$ with $V/\Gamma=1$. (b) $\Gamma \times \omega_\mathrm{F}$ with $V=\Omega$. (c) $V\times\omega_\mathrm{F}$ with $\Omega/\Gamma=10$. Red lines denote exceptional lines. }
\end{figure}

We adapt the interaction term in $\hat H_{\mathcal{PT}}$ [Eq.$~$\ref{eq3}] for a two-particle system from $V \rightarrow V \sin(\omega_\mathrm{F} t)$, and calculate the $\mathcal{PT}$ symmetry phase diagrams as the system parameters are varied.
In each diagram in Fig.~(\ref{fig12}) $\omega_\mathrm{F}$ and either $\Omega$, $\Gamma$ or $V$ are varied, while the other two parameters are fixed.
We see that modulation enriches the phase diagrams, because new $\mathcal{PT}$-broken regions appear when the unmodulated model is in the $\mathcal{PT}$-symmetric phase (see Fig.~\ref{fig12}). Note that this behavior clearly occurs in Fig.~\ref{fig12}c, where $V$ varies with $\omega_F$, showing a phase transition in the $\mathcal{PT}$-symmetric phase with $\Omega=10\Gamma$ of the unmodulated model (see Fig.~\ref{fig3}a).

In the Appendix (\ref{appendix_3}), we show the $\mathcal{PT}$-phase diagrams for systems of one and two particles when square wave modulations are applied.

\begin{figure}[t]
	\centering
	\includegraphics[height=9
		cm]{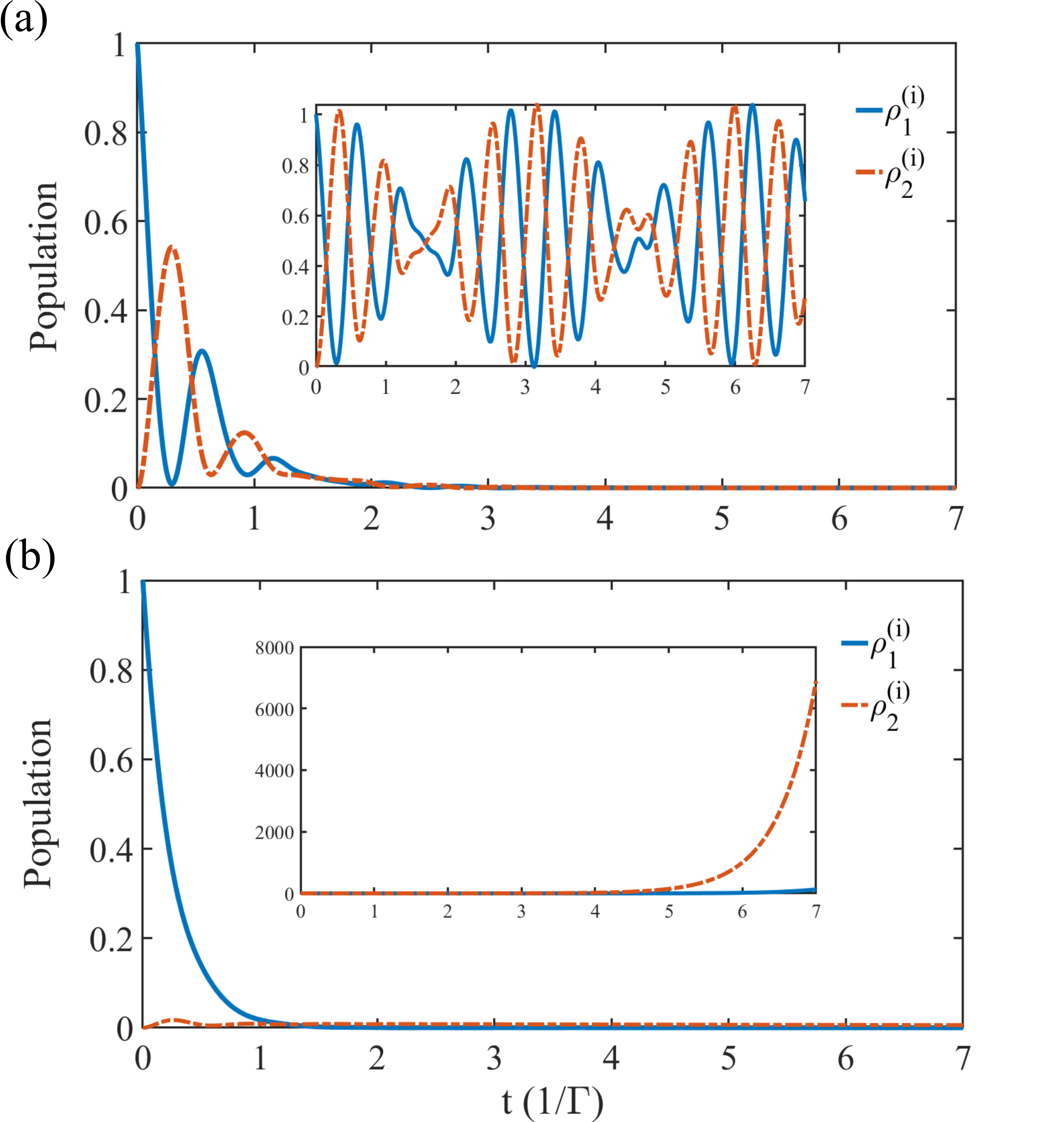}
\caption{\label{fig10} Main figures: physical populations of states $\ket{1}$ (blue line) and $\ket{2}$ (orange dotted line) for two interacting particles. (a) Population in the $\mathcal{PT}$-unbroken phase with $\Omega/\Gamma=10$ and $V/\Gamma=2$, showing the oscillatory behavior and Rabi oscillations between the two states. 
(b) Population in the $\mathcal{PT}$-broken phase with $\Omega/\Gamma=2$ and $V/\Gamma=10$. Inset: populations of $H_{\mathcal{PT}}$ in the (a) $\mathcal{PT}$-unbroken and (b) $\mathcal{PT}$-broken phases. We use $\rho^{(i)}_{1,2}$ to designate the population of the particle labeled by the index $i=1,2,3,4$ in the states $\ket{1}, \ket{2}$.} 
\end{figure}

\section{Dynamics and experimental implementation}\label{sec3}
The $\mathcal{PT}$ symmetry of a quantum system can be probed by measuring the system's dynamics \cite{Li2019}. The dynamics of the $\mathcal{PT}$-symmetric Hamiltonian is understood through the density matrix $\rho (t)=\ketbra{\psi_0 (t)}{\psi_0 (t)}$, where the state $\ket{\psi_0(t)}= \exp^{-i\hat H_{\mathcal{PT}}t}\ket{1}$. In the $\mathcal{PT}$-symmetric phase, the eigenvalues are real and the elements of the density matrix represent the population of the state with limitations on the Rabi oscillations. When the system parameters are set into the $\mathcal{PT}$-broken region the eigenvalues are complex and the population increases exponentially \cite{Wang2021,Ding2021}.

A system with effective non-Hermitian Hamiltonian $\mathcal{H}_{\rm{eff}}$ [Eq.~(\ref{eq2})] and density matrix $\hat\rho(t)$ evolves according to
\begin{eqnarray}\label{eq6}\nonumber
\frac{d\hat{\rho}(t)}{dt}&=&\frac{1}{i\hbar}\left(\hat{\mathcal{H}}_{\rm{eff}}\hat{\rho}(t)-\hat{\rho}(t)\hat{\mathcal{H}}^{\dagger}_{\rm{eff}}\right)+\Gamma_1\sum_{n=1}^{N}C_{1}^{n}\hat{\rho(t)}C_{1}^{n\dagger}\\ &&+\Gamma_2\sum_{n=1}^{N}C_{2}^{n}\hat{\rho(t)}C_{2}^{n\dagger}.
\end{eqnarray}
where $C_1 \equiv \ketbra{0}{1}$ and $C_2 \equiv \ketbra{0}{2}$ are the collapse operators of the states $\ket{1}$ and $\ket{2}$.
By measuring elements of $\hat{\rho}(t)$ (such as the populations) $
\mathcal{H}_{\rm{eff}}$ can be estimated, and therefrom $\mathcal{H}_{\mathcal{PT}}$ can be estimated and the $\mathcal{PT}$-symmetry of the system can be determined.

The $3$-level system in Fig.~(\ref{fig1}) has a dynamics described by the master equation in Eq.~(\ref{eq6}). This dynamics of the $3$-level system is connected and reduced through the dynamics of the $2$-level system by $\dot{\hat{\rho}}(t)=\frac{1}{i \hbar }\left[\hat{\mathcal{H}}_{\mathrm{eff}}\hat{\rho}(t)-\hat{\rho}(t)\hat{\mathcal{H}}_{\mathrm{eff}}^{\dagger}\right]$, and from now on we will use this form to study the dynamics of $H_{\mathcal{PT}}$. We illustrate the different system dynamics that arise for a two-particle dissipative system with a $\mathcal{PT}$-symmetric Hamiltonian and a system with a $\mathcal{PT}$-broken Hamiltonian in Fig.~(\ref{fig10}).

Note that in terms of the $\mathcal{PT}$-symmetric Hamiltonian $\hat{\mathcal{H}}_{\mathcal{PT}}$ [Eq.~(\ref{eq3})] the system evolution is described by
\begin{eqnarray}\label{eq9}
\frac{d\tilde{\rho}(t)}{dt}=\frac{1}{i \hbar}\left(\hat{\mathcal{H}}_{\mathcal{PT}}\tilde{\rho}(t)-\tilde{\rho}(t)\hat{\mathcal{H}}_{\mathcal{PT}}^{\dagger}\right).
\end{eqnarray}
The exponential population increase (see Fig.~\ref{fig10}b inset) is reversed when we rewrite the density matrix in the form
\begin{eqnarray}\label{eq7}
\hat{\rho}(t)=\tilde{\rho}(t) \ \exp(-\Gamma^{'}t),
\end{eqnarray}
where $\Gamma^{'}=\left(\Gamma_{1}+\Gamma_{2}\right)/2$. 

The density matrix $\tilde{\rho}(t)$ describes the dynamics for physical populations. Fig.$~$\ref{fig10}(a) shows an oscillatory behavior of physical populations in the $\mathcal{PT}$-symmetric region. In the inset, we also observe population oscillations for the unmodified density matrix. In Fig.$~$\ref{fig10}(b) the $\mathcal{PT}$-broken region is shown, where physical populations rapidly decay depopulating the Rydberg states. In the inset, an opposite behavior is shown, in which the population of the lower $\ket{1}$ increases exponentially.

\subsection{Experimental implementation}\label{sec3a}

Systems of Rydberg atoms and systems of Rydberg ions are well suited for implementing the Hamiltonian in Eq.~\ref{eq2}.
They can be exquisitely controlled \cite{Barredo2015,Ravets:16,Browaeys2020Many,Ebadi2021} and they exhibit strong dipolar interactions \cite{Browaeys2016Experimental}.
In a Rydberg system we envisage states $|1\rangle$ and $|2\rangle$ being Rydberg $S_{1/2}$ and $P_{1/2}$ states, with principal quantum numbers around 50.
The $XY$ interaction of the form of Eq.~\ref{eq2} can be achieved by coupling the Rydberg $S_{1/2}$ and $P_{1/2}$ states using a near-resonant microwave field; an interaction of this form was used to entangle two trapped ions \cite{Zhang2020}.
Laser light and microwave radiation are used to control systems of Rydberg atoms or ions, including preparing the particles in different states and measuring them in different bases.

By dressing the Rydberg state to other Rydberg or low lying states, the parameter values can be adjusted across large ranges, which will allow the $\mathcal{PT}$ phase diagrams of Figs.~\ref{fig3} and \ref{fig12} to be studied.
With $\mathrm{^{88}Sr^+}$ Rydberg $S_{1/2}$ and $P_{1/2}$ states with principal quantum numbers around $50$, the Rydberg states decay due to natural decay and transitions driven by blackbody radiation, with a rate $\Gamma \sim 2 \pi \times 10\,\mathrm{kHz}$.
$\Gamma$ can be tuned by changing the principal quantum number (higher states have lower decay rates), or else $\Gamma$ can be increased by adding an additional decay channel using a near-resonant laser field; for instance $\Gamma$ can easily be increased to $2\pi\times2\,\mathrm{MHz}$ by using a 306\,nm laser field which couples $nS_{1/2}  \leftrightarrow 6P_{3/2}$ in the $\mathrm{^{88}Sr^+}$ experiment. 

The coupling strength $\Omega$ between $\ket{1}$ and $\ket{2}$ (that is the Rydberg $S_{1/2}$ and $P_{1/2}$ states), can be tuned between $0$ and $2\pi\times500\,\mathrm{MHz}$ by changing the intensity of the coupling microwave field.

The interaction strength $V$ can be tuned between $0$ and $\approx 2\pi\times10\,\mathrm{MHz}$ by changing either the principal quantum number of the Rydberg states (lower states interact more weakly), by changing the distance between the ions, else by changing the detuning of the microwave field from the $S_{1/2} \leftrightarrow P_{1/2}$ resonance \cite{Zhang2020}.
Typical ion separations are around $4\,\mathrm{\mu m}$.

To modulate the system parameters (as described in Section~\ref{sec2b}) the laser light intensity or the microwave field strength can be modulated, alternatively these fields may be detuned, and the detuning may be modulated.
Modulation frequencies $\omega_\mathrm{F}$ between $0$ and $\sim 2\pi\times100\,\mathrm{MHz}$ are achievable.

Ions are routinely trapped in both linear and zigzag configurations \cite{Raizen1992}, while the configurations of atoms in dipole traps can be highly controlled \cite{Barredo2018}.
We simulated systems with interactions $V_{ij}\sim1/r_{ij}^3$, which has no angular dependence.
An interaction of this form can be achieved in a system with dipolar interactions when the particles are in a $1$D ($2$D) configuration when their dipole moments are perpendicular to the chain (plane) of the particles.

\section{Conclusions}\label{sec5}
In this work we showed that platforms with few-particles with long-range binary interactions are an excellent playground for studying $\mathcal{PT}$-symmetry phase transitions.
We find the $\mathcal{PT}$-symmetry phase diagrams become richer as the symmetry of the system is reduced.
By modulating the system parameters the phase diagrams can be further enriched.
We outline how the $\mathcal{PT}$-symmetry of a system can be determined experimentally, by measuring the evolution of the populations in different states.
We expect systems of Rydberg atoms or ions, which exhibit strong dipolar interactions, are particularly suited for experimental studies of the phenomena discussed in this work. In addition, interesting $\mathcal{PT}$-symmetry phase transitions are found by changing the Rydberg atoms configurations. This investigation generalizes previous theoretical and experimental research on effective single-particle models, to a many-body environment with direct applications to quantum simulations and quantum information processing.

\begin{acknowledgments}
We gratefully acknowledge stimulating discussions with Y. Pará.
T.M.~ acknowledges CNPq for support through 
Bolsa de produtividade em Pesquisa n.311079/2015-6. 
This work was supported by the Serrapilheira Institute 
(grant number Serra-1812-27802), by the Knut \& Alice Wallenberg Foundation (through the Wallenberg Centre for Quantum Technology [WACQT])
and by the CAPES-STINT project "Strong correlations in Cavity and Ion Quantum Simulators". 
J. L. acknowledges Stockholm University for its hospitality.
We thank the High Performance Computing Center (NPAD) at UFRN for providing computational resources.

\end{acknowledgments}
 
\section{Appendix -- interaction strengths in zigzag case}\label{appendix_1}
For three particles, the separations between the particles satisfy $r_{12}=r_{23}$ and $r_{13}=2r_{12} \sin{(\alpha/2)}$.
Thus the interactions strengths are $V_{12}=V_{13}=V$ and $V_{23}=V/(8\sin^3(\alpha/2))$.
For four particles we consider the interactions $V_{12}=V_{23}=V_{34}=\tilde{V}/r^3$, $V_{14}$ and $V_{24}$ takes the following form

\begin{eqnarray}\label{eq9z}
	V_{13} &=&\frac{V}{ 8 \ {\sin}^3 \left(\alpha_1 / 2 \right)},
\end{eqnarray}
\begin{eqnarray}\label{eq10z}
V_{24} &=&\frac{V}{ 8 \ {\sin}^3 \left(\alpha_2 / 2 \right)},
\end{eqnarray}
\begin{eqnarray}\label{eq11z}\nonumber
V_{14} &=&\frac{V}{\left(1+ 4 \ {\sin}^2 \left(\alpha_1 / 2 \right)-4 \ {\sin} \left(\alpha_1 / 2 \right)\cos\left(\alpha_2+\beta_2\right)\right)^{3/2}},\\
\end{eqnarray}
where $\alpha_1$, $\alpha_2$, $\beta_1$ and $\beta_2$ are the angles between particles. For the forms of the interactions Eq.$~$(\ref{eq9z}), Eq.$~$(\ref{eq10z}) and Eq.$~$(\ref{eq11z}), we take the distances $r_{12} = r_{23} = r_{34}= r$. Therefore, $\alpha_1 = \alpha_2 = \alpha$ and $\beta_1 = \beta_2 = \beta$, thus, we find $V_{13} = V_{24}$ and $V_{14}$ 
\begin{eqnarray}\nonumber
V_{14} &=&\frac{V}{\left(1+ 4 \ {\sin}^2 \left(\alpha / 2 \right)-4 \ {\sin} \left(\alpha / 2 \right)\cos\left(\alpha+\beta\right)\right)^{3/2}}.\\\label{eq112z}
\end{eqnarray}

\section{Appendix -- Dynamics for tree- and four particles}\label{appendix_2}

In the dynamics for the cases of three- and four- particles interactions up to next-next-nearest-neighbors were considered. 
We chose for the first, second and third neighbors $V_{\text{first}}=V$, $V_{\text{second}}=V/8$ and $V_{\text{third}}=V/27$, respectively. 

\begin{figure*}[t]
	\centering
	\includegraphics[height=10
		cm]{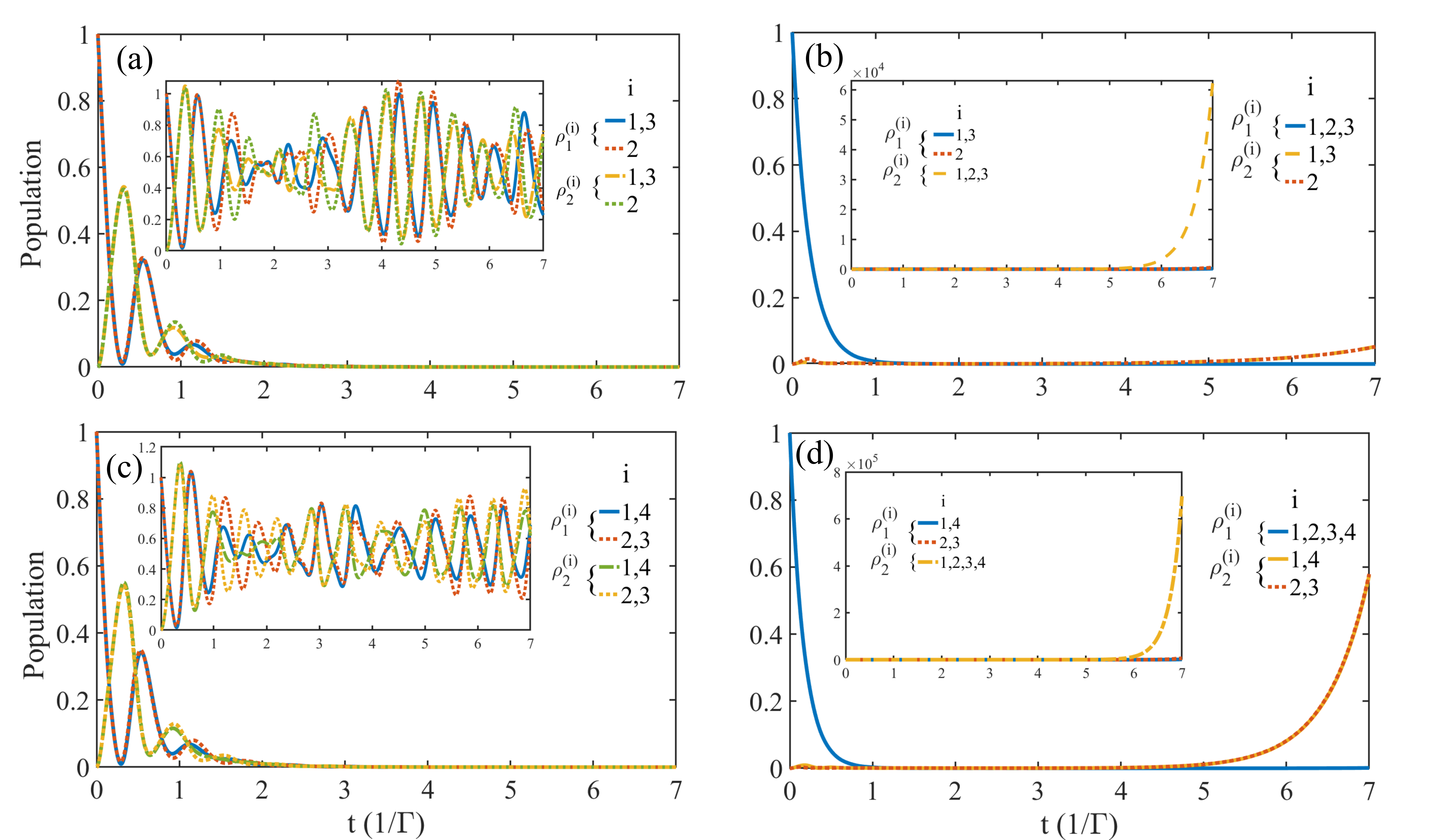}
	\caption{\label{fig13} Physical populations of Rydberg states $\ket{1}$ and $\ket{2}$ for $N=3$ and $N=4$ interacting particles. 
	(a) and (c) $\mathcal{PT}$-symmetric phase ($\Omega/\Gamma=10$ and $V/\Gamma=2$) with oscillatory behavior of Rydberg-state populations. 
	For $N=3$ particles, the populations of particle $1$ is equal to particle $3$ by symmetry in $\rho^{(i)}_{1}$ and $\rho^{(i)}_{2}$. In the $N=4$ case, the populations of particle $1$ equals the populations of particle $4$, and similarly for particles $2$ and $3$. Inset in the $\mathcal{PT}$-unbroken: oscillating populations following the dynamics according to $H_{\mathcal{PT}}$. 
	(b) and (d) phase $\mathcal{PT}$-broken ($\Omega/\Gamma=2$ and $V/\Gamma=10$) in which the physical populations decay. In the inset, in the $\mathcal{PT}$-broken phase there is an exponential increase of the populations.}
\end{figure*}

In Fig.$~$(\ref{fig13}), we show the physical populations and the populations of the unmodified density matrix for $N=3$ and $4$ particles, respectively. 
In the $\mathcal{PT}$-symmetric region, $\Omega=10\Gamma$ and $V=2\Gamma$ (see Fig.$~$\ref{fig13} ((a)-(c)), we find an oscillatory behavior of the states $\ket{1}$ and $\ket{2}$ for the physical populations. Still in the $\mathcal{PT}$-unbroken phase, in the inset the populations of the unmodified density matrix are shown, also displaying an oscillatory behavior. In the $\mathcal{PT}$-broken region (see Fig.$~$\ref{fig13} ((b)-(d))), $\Omega=2 \Gamma$ and $V=10\, \Gamma$, the physical populations of $\rho^{(3)}_{1}$ decay and are depopulated, while the population of $\rho^{(4)}_{1}$ grows smoothly.
However, in the inset of the $\mathcal{PT}$-broken phase, the populations increase exponentially, see Figs.$~$\ref{fig13}((b)-(d)).
 
 \section{Appendix -- square-wave modulation of parameters in $H_{\mathcal{PT}}$}\label{appendix_3}
\begin{figure*}[]
	\centering
	\includegraphics[height=4
		cm]{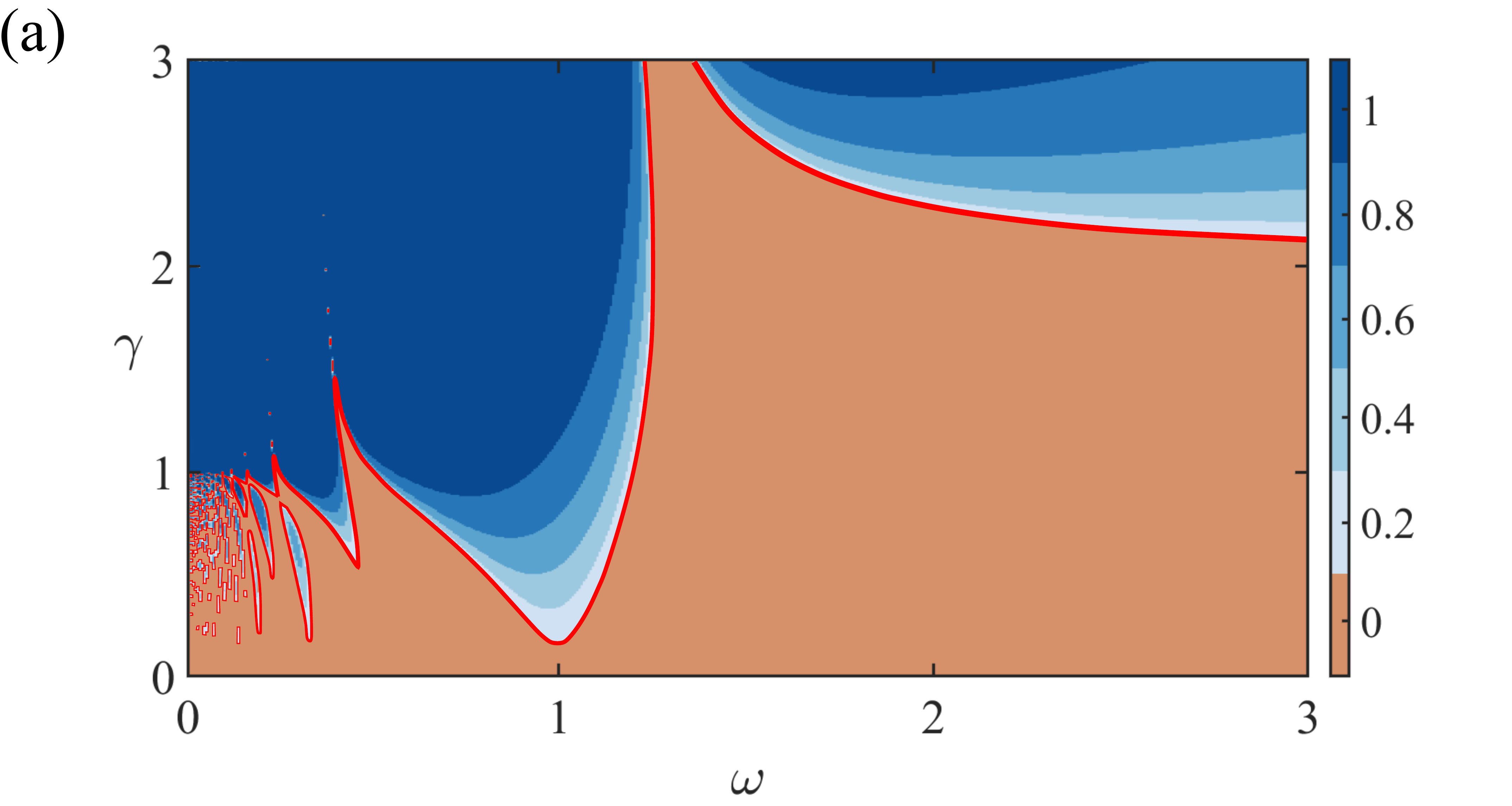}
	\includegraphics[height=4
		cm]{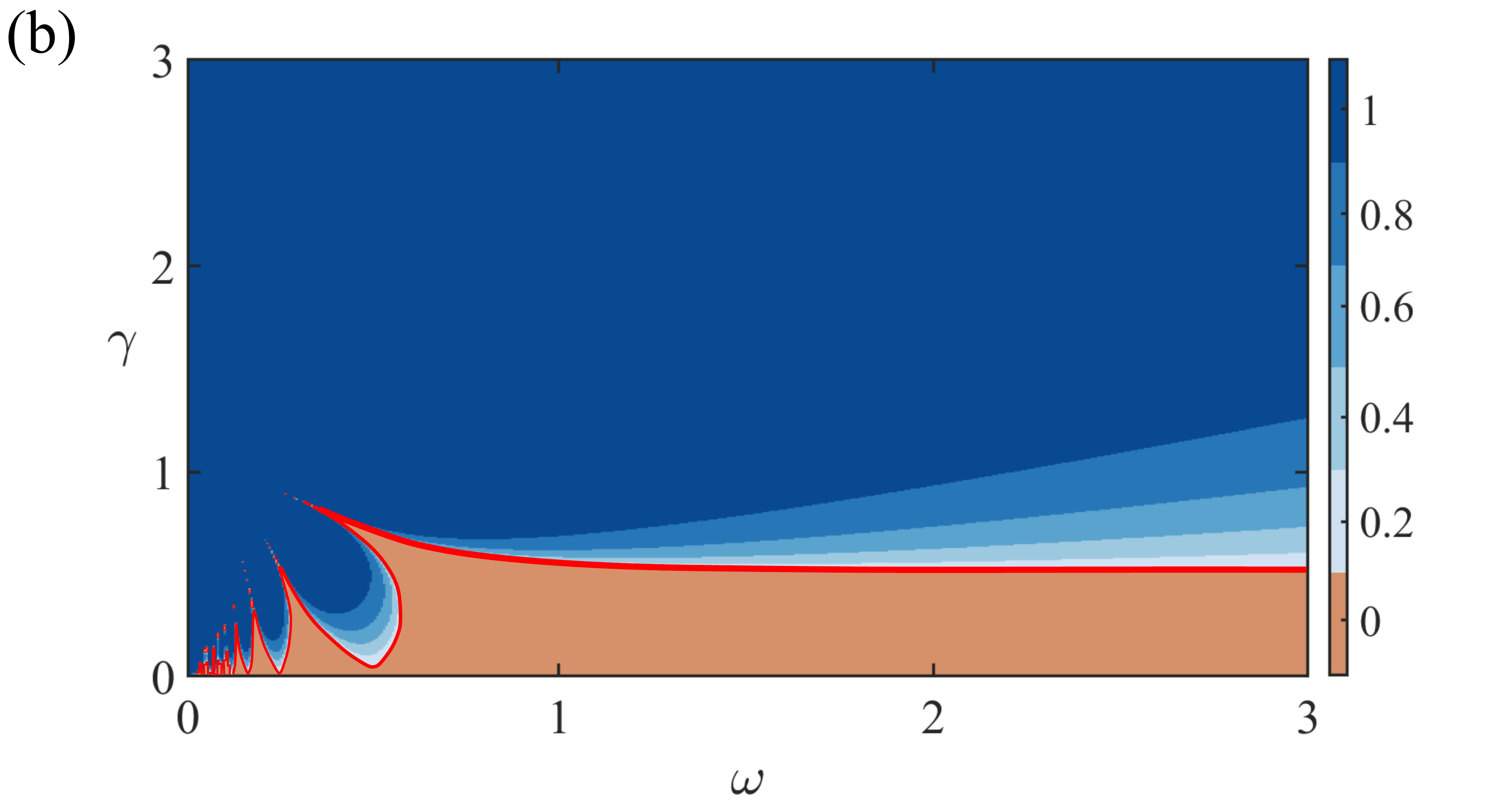}\\
	\includegraphics[height=4
		cm]{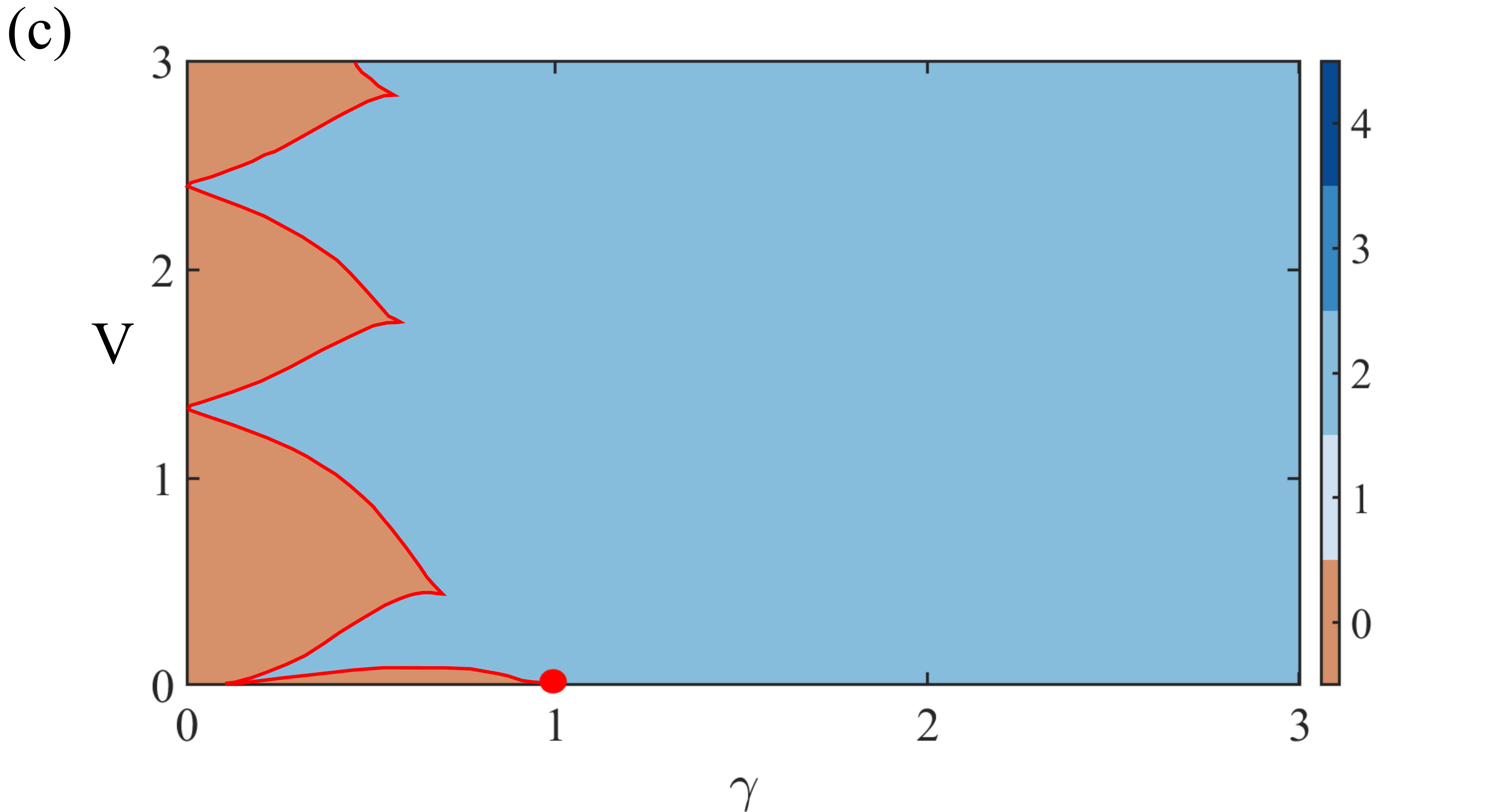} 
			\includegraphics[height=4
		cm]{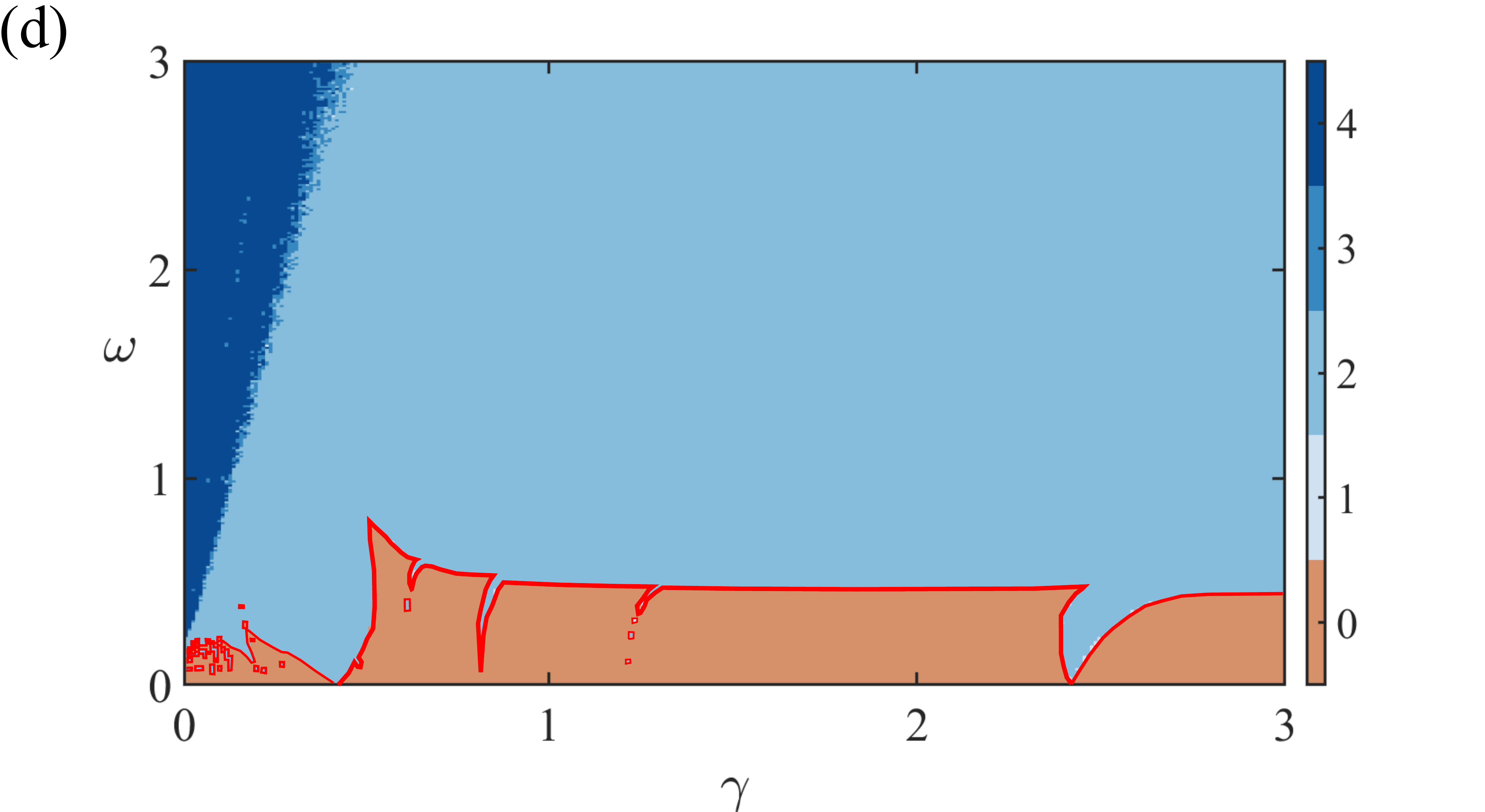} 
	\caption{\label{fig14} $\mathcal{PT}$-phase transition diagrams for Hamiltonians of one and two atoms with square-wave modulation. In the case one atom, the $\mathcal{PT}$-phase transition is represented by $\Delta e$. The $\mathcal{PT}$-unbroken phase occurs in $\Delta_e=0$ and the $\mathcal{PT}$-broken phase consists of the colored region for $\Delta e>0$. (a) and (b) $\mathcal{PT}$ transition of one atom propagator in Eq.$~$(\ref{eq34}). (c) and (d) $\mathcal{PT}$-phase transition of the propagator of two atoms in Eq.$~$(\ref{eq18}) from eigenvalues number with the module different from 1.}
\end{figure*}

The Floquet theory applied to $\mathcal{PT}$-symmetry allows the identification of $\mathcal{PT}$-broken phase transitions even in $\mathcal{PT}$-symmetric regions where the eigenvalues of the $\mathcal{PT}$-symmetric Hamiltonian are real \cite{Li2019}. The Hamiltonian of the periodically modulated system is given by
\begin{eqnarray}\label{eq24}
	H(t)=H(t+T),
\end{eqnarray}
where $T$ is the modulation period. In the case of one particle, the periodically modulated $\mathcal{PT}$-symmetric Hamiltonian is given by
\begin{equation}\label{eq25}
	\hat{H}_{\mathcal{PT}}(t)=\frac{\Omega\left(t\right)}{2}\hat{\sigma}_{x}+i\frac{\Gamma\left(t\right)}{2}\hat{\sigma}_{z}.
\end{equation}
For a single particle the propagator reads
\begin{align}\label{eq26}
	G_{\mathcal{PT}}(t) & =\text{e}^{-iH_{\mathcal{PT}}t}\nonumber \\
	& =\cos\left(E_{\pm}t\right)\mathbb{I}-i\sin\left(E_{\pm}t\right)\frac{H_{\mathcal{PT}}}{E_{\pm}},
\end{align}
where $\mathbb{I}$ is the $2\times2$ identity matrix and $E_{\pm}$ are the eigenvalues of $H_\mathcal{PT}$. The parameter that indicates the $\mathcal{PT}$-phase transition is given by
\begin{equation}\label{eq27}
	\Delta e=\frac{|e_{1}|-|e_{2}|}{|e_{1}|+|e_{2}|},
\end{equation}
where $e_{1}$ and $e_{2}$ are the eigenvalues of $G_{\mathcal{PT}}$. When $\Delta e=0$, ${\mathcal{PT}}$ symmetry is preserved, while when $\Delta e \neq 0$ ${\mathcal{PT}}$ symmetry is broken.

The coupling $\Omega$ is fixed and the dissipation $\Gamma$ is modulated between $\Gamma_0$ and $0$ with the Floquet modulation frequency $\omega_{0}$.
Using the following square-wave modulation for $0\leq t<T$, where $T=2\pi/\omega_{0}$ is the period of modulation
\begin{align}\label{eq32}
	\Gamma(t) & =\begin{cases}
		\Gamma_{0} & 0\leq t<T/2,\\
		0 & T/2\leq t\leq T,
	\end{cases}=\begin{cases}
		\Gamma_{0} & 0\leq t<\pi/\omega_{0},\\
		0 & \pi/\omega_{0}\leq t\leq2\pi/\omega_{0}.
	\end{cases}
\end{align}
The time evolution of the system over one period of the driving is the product of the propagators for the static Hamiltonian associated with each step:
\begin{equation}\label{eq33}
	G_{\mathcal{PT}}(t)=\text{e}^{-iH_{\mathcal{PT}}\left(\Gamma(t)=\Gamma_{0}\right)\tau}\text{e}^{-iH_{\mathcal{PT}}\left(\Gamma(t)=0\right)\tau},
\end{equation}
where $\tau=T/2$. Therefore, replacing the modulation conditions, we find
\begin{align}\label{eq34}
	G_{\mathcal{PT}}\left(\frac{2\pi}{\omega_{0}}\right)=\text{exp}\left[-i\left(\frac{\hat{\sigma}_{x}}{2\omega}-i\frac{\gamma}{2\omega}\hat{\sigma}_{z}\right)\pi\right]\text{exp}\left[-i\frac{\hat{\sigma}_{x}}{2\omega}\pi\right],
\end{align}
where $\omega\equiv\omega_0/\Omega$, $\gamma\equiv\Gamma_0/\Omega$ and $V\equiv V_0/\Omega$. Then, we study the $\mathcal{\mathcal{PT}}$-phase
diagram of modulated periodic dissipation $\Gamma$ and Rabi frequency $\Omega$. The modulation in $\Gamma$ is shown in Fig.$~$\ref{fig14}(a), in which the peaks are non-zero values for $\Delta e$. 
When $\Delta e$ vanishes the system is in the $\mathcal{PT}$-unbroken phase. 
Fig.$~$\ref{fig14}(b) shows the square-wave modulation in $\Omega$ similar the Eq.$~$(\ref{eq27}), where again there are $\mathcal{PT}$ transition regions. The colored region describes the $\mathcal{PT}$-broken phase, where $\Delta e>0$ and the dark blue region the $\mathcal{PT}$-unbroken phase at $\Delta e=0$.

The previous approach describes the time period for one particle. Now, we study the model of two particles in Eq.$~$(\ref{eq3}) with periodic modulation. We analyze the phase transition through the relationship of $\Delta e$ in Eq.$~$(\ref{eq27}). Note that for two particles the $\mathcal{PT}$-symmetric non-Hermitian Hamiltonian has four eigenvalues and the $\Delta e$ has a non-zero value when one of the eigenvalues has modulus different from $1$, indicating the breaking of $\mathcal{PT}$ symmetry. First, we consider the square wave modulation for the $V$ interaction similar to Eq.$~$(\ref{eq32}), and we keep $\Omega$ and $\Gamma$ fixed. The propagator then reads
\begin{widetext}
	\begin{align}\label{eq36}
		G_{\mathcal{PT}}(T)= \text{exp}\left[-i\left(\frac{1}{2\omega}\left(\hat{\sigma}_{x}^{2}+\hat{\sigma}_{x}^{1}\right)-i\frac{\gamma}{2\omega}\left(\hat{\sigma}_{z}^{2}+\hat{\sigma}_{z}^{1}\right)+\frac{V}{\omega}\left(\hat{\sigma}_{x}^{1}\hat{\sigma}_{x}^{2}+\hat{\sigma}_{y}^{1}\hat{\sigma}_{y}^{2}\right)\right)\pi\right]\text{exp}\left[-i\left(\frac{1}{2\omega}\left(\hat{\sigma}_{x}^{2}+\hat{\sigma}_{x}^{1}\right)-i\frac{\gamma}{2\omega}\left(\hat{\sigma}_{z}^{2}+\hat{\sigma}_{z}^{1}\right)\right)\pi\right].
	\end{align}
\end{widetext}

In Fig.$~$\ref{fig14}(c), we use $\omega=1$, in
this case, we have the regime in which the Rabi frequency $\Omega$ is equal to the Floquet frequency $\omega_{0}$. 
Note that in the noninteracting regime $V=0$, the $\mathcal{PT}$-symmetry is preserved for values of $\gamma<1$, and for $\gamma>1$ the $\mathcal{PT}$-symmetry is broken, presenting an exceptional point in $\gamma=1$ described by the red dot, which in this static case also presents the same exceptional point. The exceptional line is given by the red line that separates the $\mathcal{PT}$-symmetric and broken regions.  

We take the modulation for the $\Gamma$ dissipation, whose propagator is given through square wave modulation in the form of Eq.$~$(\ref{eq32}).Setting $V=\Gamma$, we find the $\mathcal{PT}$-phase transition diagram, which also has symmetry breaking phases (see Fig.$~$\ref{fig14}(d)).
 
\section{Details on the mapping of eq.(\ref{eq7})}
\subsection{System population with $\mathcal{PT}$ dynamics from the experimental $\rho_{00}$}
Consider the master equation for the effective Hamiltonian (\ref{eq2}) written as follows:

\begin{eqnarray}\label{eq6.1}\nonumber
\frac{d\hat{\rho}(t)}{dt}&=&\frac{1}{i}\left[\hat{\mathcal{H}}_{\rm{eff}}\hat{\rho}(t)-\hat{\rho}(t)\hat{\mathcal{H}}^{\dagger}_{\rm{eff}}\right]\\\nonumber
&=&\frac{1}{i}\left[-i\left(\frac{\Gamma_{1}+\Gamma_{2}}{2}\right)\hat{\rho}(t)+ \hat{\mathcal{H}}_{\mathcal{PT}}\hat{\rho}(t)-\hat{\rho}(t)\hat{\mathcal{H}}_{\mathcal{PT}}^{\dagger}\right],\\
\end{eqnarray}
where we use $\hbar = 1$. For the effective Hamiltonian in Eq.$~$(\ref{eq2}). We can choose the density matrix in the form below:
\begin{eqnarray}
\hat{\rho}(t)=\tilde{\rho}(t) \ \exp\left(-\Gamma^{'}t\right),
\end{eqnarray}
and we find
\begin{eqnarray}\label{eq8}\nonumber
\frac{d\hat{\rho}(t)}{dt}=\exp\left(-\Gamma^{'}t\right)\left(\frac{d\tilde{\rho}(t)}{dt}-\Gamma^{'}\tilde{\rho}(t)\right),\\
\end{eqnarray}
where
\begin{eqnarray}\label{eq9.1}
\frac{d\tilde{\rho}(t)}{dt}=\frac{1}{i}\left(\hat{\mathcal{H}}_{\mathcal{PT}}\tilde{\rho}(t)-\tilde{\rho}(t)\hat{\mathcal{H}}_{\mathcal{PT}}^{\dagger}\right).
\end{eqnarray}
This result can be verified using 
\begin{eqnarray}\label{eq10}
\tilde{\rho}(t)=\exp\left(-i\hat{\mathcal{H}}_{\mathcal{PT}}\right)\tilde{\rho}(0) \ \exp\left(i\hat{\mathcal{H}}_{\mathcal{PT}}^{\dagger}\right),
\end{eqnarray}
with $\tilde{\rho}(0)=\hat{\rho}(0)$.
Consider the master equation Eq.$~$(\ref{eq9.1}) written as follows
\begin{eqnarray}\label{eq11}
\frac{d\tilde{\rho}(t)}{dt}=i\frac{\Omega}{2}\left(\tilde{\rho}\hat{\sigma}_x-\hat{\sigma}_x\tilde{\rho}\right)-\frac{\Gamma}{2}\left(\hat{\sigma}_z\tilde{\rho}+\tilde{\rho}\hat{\sigma}_z \right).
\end{eqnarray}
Using the master equation Eq.$~$(\ref{eq11}), we determine the optical Bloch equations of $\mathcal{PT}$-symmetric non-Hermitian Hamiltonian for $\tilde{\rho}(t)$ 

\begin{eqnarray}\label{eq12-15}
\frac{d\tilde{\rho}_{22} (t)}{dt}&=&i\frac{\Omega}{2}\left(\tilde{\rho}_{21}-\tilde{\rho}_{12}\right)-\Gamma\tilde{\rho}_{22},\\
\frac{d\tilde{\rho}_{11} (t)}{dt}&=&i\frac{\Omega}{2}\left(\tilde{\rho}_{12}-\tilde{\rho}_{21}\right)+\Gamma\tilde{\rho}_{11},\\
\frac{d\tilde{\rho}_{21} (t)}{dt}&=&i\frac{\Omega}{2}\left(\tilde{\rho}_{22}-\tilde{\rho}_{11}\right),\\
\frac{d\tilde{\rho}_{12} (t)}{dt}&=&i\frac{\Omega}{2}\left(\tilde{\rho}_{11}-\tilde{\rho}_{22}\right).
\end{eqnarray}
In this section, we seek to calculate the population of the system with $\mathcal{PT}$ dynamics for an experimental density matrix $\rho_{00}$ defined below

\begin{eqnarray}\label{eq16}\nonumber
\rho_{00}&=& 1-\exp(-\Gamma' t)\tilde{\rho}_{00}\\
&=&1-\rho_{22}-\rho_{11},
\end{eqnarray}
and $\rho_{22}$ and $\rho_{11}$ are defined 
\begin{eqnarray}\label{eq17}
\rho_{22}&=& \exp(-\Gamma' t)\tilde{\rho}_{22},\\\label{eq18}
\rho_{11}&=& \exp(-\Gamma' t)\tilde{\rho}_{11},
\end{eqnarray}
thus, 
\begin{eqnarray}\label{eq19}
\nonumber\tilde{\rho}_{00}&=& \exp(\Gamma' t)\left({\rho}_{22}+{\rho}_{00}\right)\\
&=& \tilde{\rho}_{22} + \tilde{\rho}_{11}.
\end{eqnarray}
Then, we find the time evolution of $\rho_{00}$ and $\tilde{\rho}_{00}$ given as follows
\begin{eqnarray}\label{eq20}
\frac{d{\rho}_{00}(t)}{dt}&=&\exp\left(-\Gamma' t\right)\left(\Gamma_{2}\tilde{\rho}_{22}+\Gamma_{1}\tilde{\rho}_{11}\right)\\\label{eq21}
\frac{d\tilde{\rho}_{00}(t)}{dt}&=& \Gamma\left(\tilde{\rho}_{22}-\tilde{\rho}_{11}\right).
\end{eqnarray}
Therefore, accessing both $\tilde{\rho}_{00}$ and $d\tilde{\rho}_{00}/dt$, we find
\begin{eqnarray}\label{eq22}
\tilde{\rho}_{22}=\frac{1}{2}\left(\tilde{\rho}_{00}+\frac{1}{\Gamma}\frac{d\tilde{\rho}_{00}}{dt}\right),
\end{eqnarray}
and
\begin{eqnarray}\label{eq23}
\tilde{\rho}_{11}=\frac{1}{2}\left(\tilde{\rho}_{00}-\frac{1}{\Gamma}\frac{d\tilde{\rho}_{00}}{dt}\right).
\end{eqnarray}
Therefore, the quantities in Eq.$ ~ $(\ref{eq22}) and in Eq.$~$(\ref{eq23}) are numerically calculated from the $\mathcal{PT}$-symmetric non-Hermitian Hamiltonian in Eq.$~$(\ref{eq3}).

\bibliographystyle{apsrev4-1}
\bibliography{Reference}

\end{document}